\documentclass[12pt,a4paper]{article}
\usepackage[T2A]{fontenc}
\usepackage[cp1251]{inputenc}
\usepackage{array,amsmath,amssymb,amstext,textcomp}
\usepackage{graphicx}
\usepackage{natbib}
\usepackage[russian,english]{babel}
\usepackage{wrapfig,color}
\usepackage[centerlast,small]{caption2}
\renewcommand\baselinestretch{1.57}

   \reversemarginpar\parskip=2pt plus 2pt
\begin{document}
\title{\textbf{EVIDENCE FOR STRONG CO-EVOLUTION OF MITOCHONDRIAL AND SOMATIC GENOMES}}
\renewcommand\baselinestretch{1.07}
\author{Michael G.~Sadovsky\\
Institute of computational modelling of SB RAS\\ 660036 Russia, Krasnoyarsk; \textsl{msad@icm.krasn.ru}
}
\date{\empty}
\maketitle


\newcounter{nmiton}
\renewcommand\baselinestretch{1.17}\normalsize
\begin{abstract}
We studied a relations between the triplet frequency composition of mitochondria genomes, and the phylogeny of their bearers. First, the clusters in 63dimensional space were developed due to $K$-means. Second, the clade composition of those clusters has been studied. It was found that genomes are distributed among the clusters very regularly, with strong correlation to taxonomy. Strong co-evolution manifests through this correlation: the proximity in frequency space was determined over the mitochondrion genomes, while the proximity in taxonomy was determined morphologically.
\end{abstract}\vspace{1.5cm}

\renewcommand\baselinestretch{1.39}\normalsize
\setcounter{section}{0}
\section{Introduction}
A study of statistical properties of nucleotide sequences still may tell a lot to a researcher on the relation between structure and some biological issues encoded in these former. A frequency dictionary of a nucleotide sequence is supposed to be a structure, further \citep{sadov1,sadov2,sadov3,g4}. A consistent and comprehensive study of frequency dictionaries answers the questions concerning the statistical and information properties of DNA sequences. A frequency dictionary, whatever one understands for it, is rather multidimensional entity.
%
%
Relation between the structure (i.\,e., oligonucleotides composition and their frequency), and the taxonomy of the bearers of DNA sequences is of great importance. Here we studied this relation for the set of mitochondrion genomes. 

To begin with, let's introduce basic definition. Consider a continuous symbol sequence from four-letter alphabet $\aleph = \{\mathsf{A}, \mathsf{C}, \mathsf{G}, \mathsf{T}\}$ of the length $N$; the length here is just the total number of symbols in a sequence. This sequence corresponds to some genetic entity (genome, chromosome, etc.). No other symbols or gaps in the sequence take place, by supposition (see details in Section~\ref{metody}). Any coherent string $\omega = \nu_1\nu_2\ldots \nu_q$ of the length $q$ makes a word. A set of all the words occurred within a sequence yields the support of that latter. Counting the numbers of copies $n_{\omega}$ of the words, one gets a finite dictionary; changing the numbers for the frequency \[f_{\omega} = \frac{n_{\omega}}{N}\] one gets the frequency dictionary $W_q$ of the thickness $q$. This is the main object of our study. Again, more technical details could be found in Section~\ref{metody}

Further, we shall concentrate on frequency dictionaries $W_3$ (i.\,e., the triplet composition) only. Thus, any genetic entity\footnote{That is mitochondrion genome, in our case.} is represented by a point in 63-dimensional space. Obviously, two genetic entities with identical frequency dictionaries $W^{(1)}_3$ and $W^{(2)}_3$ are mapped into the same point in the space. It is evident, that the absolute congruency of two frequency dictionaries $W^{(1)}_3$ and $W^{(2)}_3$ does not mean a complete coincidence of the original sequences standing behind the dictionaries. Nonetheless, such two sequences are indistinguishable from the point of view of their triplet composition.

Definitely, few entities may have very proximal frequencies of all the triplets, but few others may have not, thus making a distribution of the points in 63-dimensional space inhomogeneous. So, the key question here is what is the pattern of this distribution of mitochondrion genomes in that space? Are there some discrete clusters, and if yes is there a correlation to a phylogeny of the genome bearers and clusters?

To address the questions, we have implemented an unsupervised classification of mitochondrion genomes, in (metric) space of frequencies of triplets. Then, the taxa composition of the classes developed due to the classification has been studied; a considerable correlation between taxa composition, and the class occupation was found. Some results of the study of the correlation of the distribution of bacterial taxa in the information value space, developed over 16S\,RNA are presented in \citep{sadov1,16S-ASME,16S-opensyst}.

This paper presents the evidences of the strong relation between the structure of mitochondrion genomes, and the taxonomy of their bearers.

\section{Materials and methods}\label{metody}
We used the sequences from EMBL--bank (release of March, 2011). This release contains~$\sim 3.4 \times 10^3$ entries. The final database used for our studies enlists~1132 entries. \begin{wraptable}{l}{80mm}
\begin{tabular}{|l|r|l|r|}\hline
Order & $M$&Order & $M$\\\hline
\textit{Batrachia} & 51& \textit{Chondrostei} & 5\\ \textit{Crocodylidae} &7& \textit{Cryptodira} &25\\ \textit{Dinosauria} &94& \textit{Eutheria} &193\\ \textit{Gymnophiona} &16& \textit{Metatheria} &18\\ \textit{Neopterygii} &500& \textit{Squamata} &78\\\hline
\end{tabular}
\caption{\label{T1}Database structure; $M$ is the abundance of taxon.}
\end{wraptable}
This discrimination comes from the (not obvious) constraint: we had to keep in the database the entries which do not present rather highly ranked clades solely. In simple words, a taxon of order rank (or higher) presented with a single genome yields ``signal'' that is strong enough to deteriorate a general pattern, but fails to produce a distinguishable detail in the pattern. Thus, we enlist into the final database the entries representing an order with five species or more. Some genomes have other than indicated above symbols in the sequence; we have omitted those ``junk'' symbols with concatination of the sequence fragments separated by those junk symbols into a coherent entity.

The stricture of the database is following: it contains the genomes of animals, only, with~988 entries of \textit{Chordata} and~144 entries of \textit{Arthropoda} phyla. Further discretion in \textit{Chordata} phylum is shown in Table~\ref{T1}. \textit{Arthropoda} phylum consists of~83 entries of \textit{Endopterygota}, 30 entries of \textit{Paraneoptera} and 30 entries of \textit{Orthopteroidea}.

Unsupervised classification by $K$-means has been implemented to develop classes (see details and a lot of extensions in \citep{gusev,g1,g2,dyn1,dyn2}). We sequentially developed the classifications with two, three and four classes. No class separability has been checked. To develop a classification, we reduced the dimension of data to 63: this reduction comes from the fact that the sum of all frequencies must be equal to~1. Formally speaking, any triplet could be excluded from the data set; practically, we excluded triplet $\mathsf{GCG}$, since that latter yields the least standard deviation among other triplet ($\sigma_{\mathsf{GCG}} = 0.001399$). In $K$-mean applications, Euclidean distance has been used. All the results were obtained with \textsl{ViDaExpert} software by A.~Zinovyev\footnote{http://bioinfo-out.curie.fr/projects/vidaexpert/}.

\section{Results}\label{inftz}
Here we present the results of the classification implementation through $K$-means technique. Consider a classification with two classes, firstly. This classification carried out with $K$-means is very stable and very discretional: there are no volatile genomes under this classification. We have carried out $10^3$ runs of the classification, and in 13 runs a class consisting of a single element has been observed. All other runs yielded very stably the classification with two classes enlisting~154 and~978 entries, respectively. The composition of classes is following: 142 genomes of \textit{Arthropoda} always form a class, and only two genomes belong to the opposite class. These two genomes belong to \textit{Reticulitermes flavipes} and \textit{Gampsocleis gratiosa} (accession numbers EF206314 and EU527333, respectively).

Table~\ref{T2} shows the results of this classification implementation. Turtles occupy the same class, with no exclusions. Also, this class is occupied by fossils (\textit{Archosauria} and \textit{Lepidosauria}). Three clades (\textit{Neoptera} division, mammalia and fossils) are distributed \begin{wraptable}{l}{103mm} \begin{tabular}{|p{55mm}|r|r|r|r|}\hline
\multicolumn{1}{|c|}{Taxon} & \multicolumn{1}{c|}{$N$} & \multicolumn{1}{c|}{I} & \multicolumn{1}{c|}{II} & \multicolumn{1}{c|}{III}\\\hline
\textit{Actinopterygii} & 510 & 464 & 46 & 0\\
\textit{Amphibia} & 65 & 40 & 17 & 8\\
\textit{Archosauria} and \textit{Lepidosauria} & 177 & 1 & 176 & 0\\
\textit{Mammalia} & 212 & 0 & 1 & 211\\
\textit{Neoptera} & 143 & 0 & 4 & 139\\
\textit{Testudines} & 25 & 0 & 25 & 0\\\hline
\end{tabular}
\caption{\label{T2}Distribution of clades for the unsupervised classification implemented for three classes; $N$ is the abundance of a taxonomy group.}
\end{wraptable} between two classes, only, and the distribution is extremely biased: there are two single genomes of mammalia and fossils, respectively, belonging to the opposite class. Here turtles and fossils occupy the came class, that is not a matter of surprise. Less clear is an amalgamation of two rather distinct clades (these are \textit{Mammalia} and \textit{Neoptera}) into the single class. A rate of escaped genomes here is less than 0.5\,\% and 1.5\,\%, respectively.

976 genomes from \textit{Chordata} phylum form another class, with~12 genomes escaping the opposite class. Again, the escaping genomes set is absolutely stable and includes the following species (accession numbers are in parentheses): \textit{Ranodon sibiricus} (AJ419960), \textit{Aneides flavipunctatus} (AY728214), \textit{Ensatina eschscholtzii} (AY728216), \textit{Rhyacotriton variegatus} (AY728219), \textit{Desmognathus fuscus} (AY728227),
\textit{Hydromantes brunus}	(AY728234), \textit{Geotrypetes seraphini} (AY954505), \textit{Pachyhynobius shangchengensis} (DQ333812), \textit{Onychodactylus fischeri} (DQ333820), \textit{Dermophis mexicanus} (GQ244467), \textit{Dicamptodon aterrimus} (GQ368657), \textit{Hemiechinus auritus} (AB099481).

Classification implementation in three classes with $K$-means also was very good and stable. Again, $10^3$ runs of the classification development have been carried out. Only three patterns of the classification have been observed differing in the abundances of the classes. Namely, the abundance distributions were the following:
\begin{list}{(\roman{nmiton})}{\usecounter{nmiton}\leftmargin=15mm \labelwidth=5mm \topsep=0mm \labelsep=2mm \itemsep=1pt \parsep=0mm \itemindent=1pt\rightmargin=6cm}
\item $975 \div 147 \div 10$ entries, \hfill 18 cases,
\item $510\div 147 \div 475$ entries, \hfill 854 cases,
\item $511\div 146 \div 475$ entries, \hfill 136 cases.
\end{list}
Again, there were no volatile genomes (i.\,e. those permanently changing the class attribution).

Fishes tend to occupy two classes, in more explicit manner. The majority of the genomes of this clade occupy the first class (together with \textit{Amphibia}), while 10\,\% of the genomes of fishes are located at the second class. This ratio differs from the similar one observed for other clades (except \textit{Amphibia}).

\textit{Amphibia} exhibit the most intriguing behaviour. That is the only clade occupying all three classes rather remarkably. Moreover, the distribution in triplet frequency space is sensitive at quite low taxonomy level. \textit{Amphibia} enrolled~9 genomes of \textit{Caudata} order, and~13 genomes of \textit{Anura} order. These two orders are separated into two classes of the statistical classification: all \textit{Anura} genomes fall into the second class, while the third class includes~7 genomes of \textit{Caudata} order, in comparison to two genomes of that latter order found in the second class.

\begin{figure}[!h]
\centerline{\includegraphics[bb=0 0 627 624, viewport=5 3 622 620, clip=true, scale=0.69]{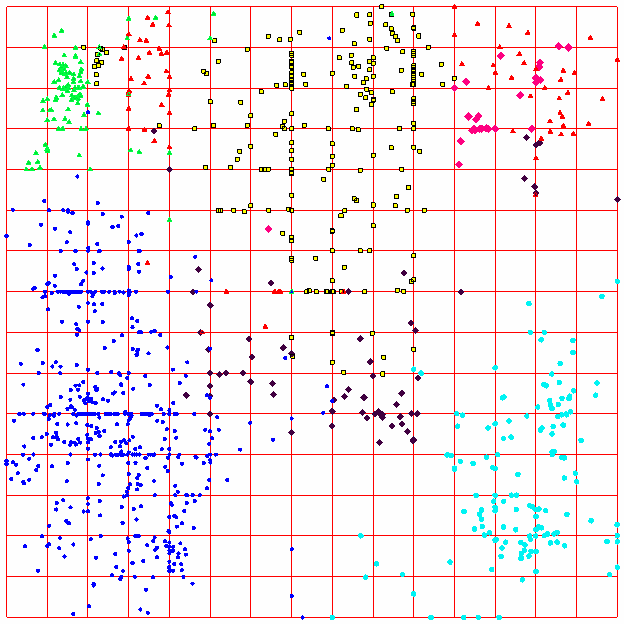}}
\caption{\label{fig1}Distribution of genomes in rigid map. Following are the labels of the clades: \mbox{\textcolor{blue}{$\bullet$}  --  \textit{Actinopterigii};} \textcolor{black}{$\blacklozenge$}  --  \textit{Amphibia}; \textcolor{green}{$\blacktriangle$} --  \textit{Archosaura}; \textcolor{red}{$\blacktriangle$} --  \textit{Lepidozaura}; \textcolor{yellow}{$\blacksquare$}\hspace{-9pt}\textcolor{black}{$\square$} --  \textit{Mammalian}; \textcolor{cyan}{$\bullet$}  --  \textit{Neoptera}; \textcolor{magenta}{$\blacklozenge$} \hspace{-11pt}\textcolor{black}{$\Diamond$} --  \textit{Testudines}.}
\end{figure}

\begin{figure}[!h]
\includegraphics[bb=0 0 627 624, viewport=5 3 622 620, clip=true, scale=0.36]{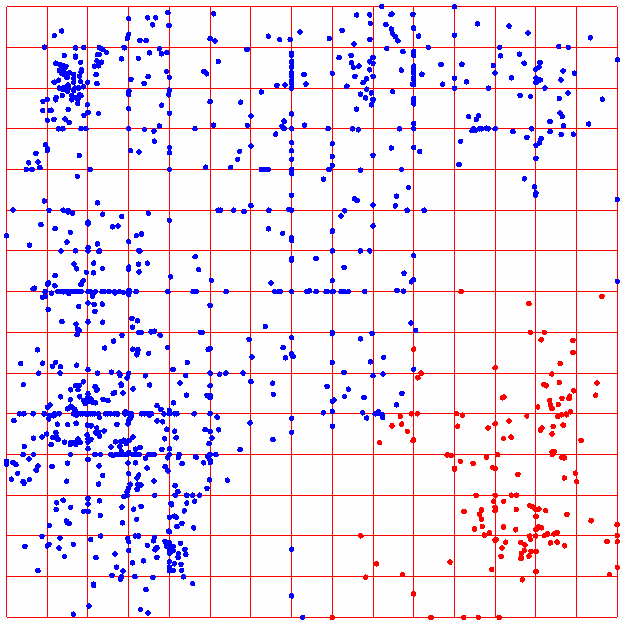}\hfill\includegraphics[bb=0 0 627 624, viewport=5 3 620 620, clip=true, scale=0.36]{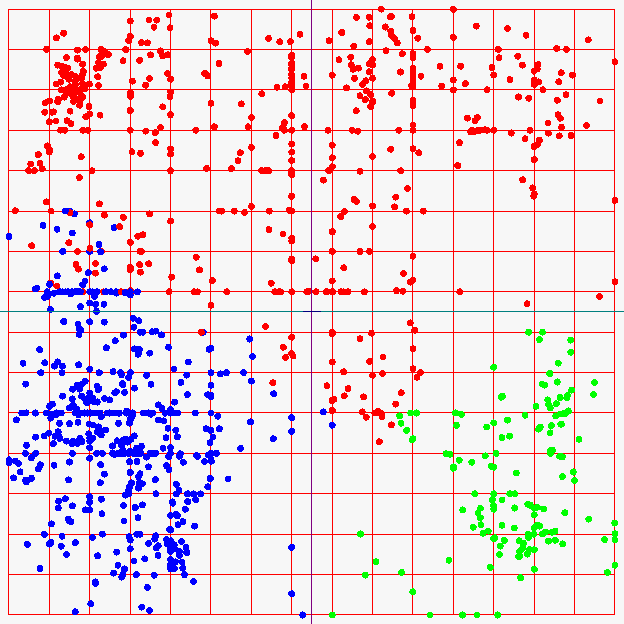}
\caption{\label{fig2}Two class (left) and three class (right) distributions of the genomes.}
\end{figure}

Fig.~\ref{fig1} illustrates the distribution of clades in rigid map (see \citep{g1,gusev} for the details on the rigid map techniques). Labels for clades are in the figure legend. Fig.~\ref{fig2} shows the distributions of the genomes for three- and two-class classification.

\section{Discussion}
A study presented in this paper is done within the scope of population genomics methodology. An idea to figure out the relation between taxonomy and triplet composition of some genetic entities was originally provided by \citep{16S-opensyst,obbiol2003}. These papers presented the study of a relation between bacterial taxonomy, and triplet frequency dictionaries determined over 16S~RNAs of bacteria. Here we apply those ideas to the study of \textsl{structure--taxonomy} relations observed over the entire genetic object (that is a mitochondrion genome).\medskip

Mitochondria are another very good objects for the population studies of this type: they have rather simple and short genome (typical length is about~$10^5$ nucleotides); they encode absolutely the same function, in any organism; a genome consists of a single chromosome, thus making taxonomy the only factor affecting the difference among them. Another good genetic system that might be used for this kind of studies are chloroplasts. Moreover, one might want to study the mutual distribution of two genomic systems: the former is of mitochondria, and the latter is of chloroplasts.

Probably, a database structure is the key problem in this kind of studies. We have used an unsupervised classification technique to develop a distribution of genomes into few groups. The results of such classification are usually quite sensitive to an original database composition \citep{dyn1,dyn2}. We have found that database containing too many entries representing higher taxa with a single (or, maybe, two) species shows nothing in the terms of a distribution of genomes into the classes, even with neither respect to the specific composition of the classes. In other words, no separation into classes takes place, for such databases.

Increasing ``concentration'' of relatively proximal species in the database, one can figure out various patterns in the set of genetic entities. In such capacity, the best database should enlist the equal number of entries in any genus. Should the number of genera be equal in the taxonomy ranks of higher order in such databases, is a matter question. Indeed, the results presented above unambiguously prove the efficiency of unsupervised linear classification technique for rather non-equilibrium databases.\medskip

Another option to retrieve more knowledge towards the relation between structure and taxonomy (or function) is to change for some other type of structure. Indeed, here we used the most general structure that was frequency dictionary~$W_3$ of triplets. Here each nucleotide generates a triplet (for the sequence connected into a ring), and the frame shift step is equal to one nucleotide. There could be other dictionaries; first of all, obtained due to a variation of the frame shift step. In particular, paper \citep{sim2} shows the results in bacterial genomes clusterization obtained through the comparison of three different dictionaries~$\widetilde{W}^{(1)}_3$, $\widetilde{W}^{(2)}_3$ and $\widetilde{W}^{(3)}_3$ counted over non-overlapping triplets, for three different starting points of a count. Obviously, a unification of these three dictionaries yields the standard dictionary~$W_3$.

Evidently, a set of frequency dictionaries (of triplets) is mapped into a linear subspace of co-dimension one. All the points representing the genomes are located at the simplex determined by the normalization constraint \[f_{\mathsf{A}}+f_{\mathsf{C}}+f_{\mathsf{G}}+f_{\mathsf{T}}=1\,.\] This constraint makes the interlocation of the points quite bound. Changing real frequencies of triplets for the information values of these latter, one can break through the constraint mentioned above. Information value of a triplet here is the ratio of real frequency $f_{\nu_1\nu_2\nu_3}$ and the expected one $\widetilde{f}_{\nu_1\nu_2\nu_3}$. Obviously, an expected frequency could be determined in various ways; the approach based on the construction of the most probable continuation of dinucleotides in triplets yields the formula \begin{equation}\label{eq:1}
\widetilde{f}_{\nu_1\nu_2\nu_3} = \dfrac{f_{\nu_1\nu_2}\times f_{\nu_2\nu_3}}{f_{\nu_2}}
\end{equation}
for the expected frequency $\widetilde{f}_{\nu_1\nu_2\nu_3}$. The details of this idea could be found in \citep{osid98,16S-ASME,sadov1,sadov2,sadov3,g4}. Thus, the information value of triplet is the ratio of real frequency and the expected determined by~\eqref{eq:1}. This approach was very fruitful, for a study of the correlation between the structure and taxonomy of bacteria \citep{16S-opensyst,obbiol2003}.

\end{document}